\documentclass[aps,pra,twocolumn,showpacs,preprintnumbers,amsmath,amssymb,footinbib]{revtex4}
\usepackage{mathtools}

\DeclareMathAlphabet{\mathcalligra}{T1}{calligra}{m}{n}
\DeclareMathAlphabet{\mathpzc}{OT1}{pzc}{m}{it}

\usepackage{graphicx,epsfig}
\usepackage{bm}
\usepackage{dcolumn}
\usepackage{xcolor}
\usepackage[breaklinks=true,colorlinks,citecolor=blue,linkcolor=blue,urlcolor=blue]{hyperref}

\def\sc#1#2#3{Science {\bf #1}, #2 (#3)}
\def\prv#1#2#3{Phys. Rev. {\bf #1}, #2 (#3)}

\def\prl#1#2#3{Phys. Rev. Lett. {\bf #1}, #2 (#3)}
\def\pra#1#2#3{Phys. Rev. A {\bf #1}, #2 (#3)}

\def\noi{\noindent}
\def\bc{\begin{center}}
\def\ec{\end{center}}
\newcommand{\bea}{\begin{equation}}
\newcommand{\eea}{\end{equation}\noi}
\newcommand{\ber}{\begin{eqnarray}}
\newcommand{\eer}{\end{eqnarray}\noi}
\begin{document}
\title{Refractive index for the mechanical refraction of a relativistic particle}

\author{Bikram Keshari Behera}
\author{Surendra Kumar Gour\footnote{Present address: Department of Physics, Indian Institute of Technology Guwahati, Assam-781039, India}}
\author{Shyamal Biswas}\email{sbsp [at] uohyd.ac.in}

\affiliation{School of Physics, University of Hyderabad, C.R. Rao Road, Gachibowli, Hyderabad-500046, India
}

\date{\today}

\begin{abstract}
We have analytically determined the refractive index for the mechanical refraction of a relativistic particle for its all possible speeds. We have critically analysed the importance of Descartes' metaphysical theory and extended it in this regard. We have considered the conservation of the tangential component of the relativistic momentum and the relativistic energy of the particle in the process of the mechanical refraction within the optical-mechanical analogy. Our result for the mechanical refractive index exactly matches with the forms of both the Fermat's result on Snell's law of optical refraction at the ultra-relativistic limit and the Descartes' metaphysical result on the pseudo-Snell law of optical refraction at the non-relativistic limit. 

\end{abstract}

\pacs{42.25.Gy (Edge and boundary effects; reflection and refraction), 01.55.+b (General physics), 45.20.dh (Energy conservation), 03.30.+p (Special relativity)}


\maketitle    

\section{Introduction}

Historically, the law of refraction of light was discovered from the experimental observations by Ibn Sahl in the year 984 \cite{Snell1} and Snellius in the year 1621 \cite{Snell2}. The law of optical refraction, however, is known as Snell's law: $\frac{\sin(\theta_i)}{\sin(\theta_r)}=n_{ri}$, according to which - the ratio of the sine of the angle of incidence ($\theta_i$) to the sine of the angle of refraction  ($\theta_r$) is a non-universal constant ($n_{ri}$) which depends on the optical properties of the two media \cite{Snell2}. Later the non-universal constant $n_{ri}$ was equated with the ratio of the speed of light in the medium of incidence ($c_i$) to that in the refracting medium ($c_r$) by Huygens within the geometrical optics limit\footnote{In the limiting case of the geometrical optics the wavelength ($\lambda$) of a monochromatic wave in an optical medium is so small compared to the lowest dimension ($d$) of an aperture in the medium that it satisfies $\lambda/d\rightarrow0$.} of Huygens-Fresnel principle in 1678 \cite{Huygens,Fresnel} soon after Rømer's discovery of the finiteness of the speed of light \cite{Roemer}. The phase speed of light, however, was not known this time and the speed of light had been treated like the speed of a particle even before and after this work. Later in 1704 Newton defined the non-universal constant ($n_{ri}$) as the relative refractive index of the refracting medium with respect to the medium of incidence in connection with the colour theory and the corpuscular theory \cite{Newton}. The refractive index of an optical medium ($n$) was defined much later in 1807 by Young as a ratio of the speed of light in free space ($c$) to the (phase) speed of light in the refracting medium ($c_r$) \cite{Young}. The wave theory of light significantly progressed at around and after this time especially after Fresnel's 1818's work on the diffraction of light \cite{Fresnel}. The phase (angle) and the phase velocity of light were realized by Fresnel in connection with this work. Since then the speed of light in a medium had been treated as its phase speed \cite{Fresnel2} until its distinction from the group speed was made by Hamilton in 1839 \cite{Hamilton,Rayleigh}. The modern form of Snell's law is expressed as $n_r\sin(\theta_r)=n_i\sin(\theta_i)$ where $n_i$ is the refractive index of the medium of incidence and $n_r$ is the refractive index of the refracting medium which are expressed in terms of the permittivities and permeabilities of the respective optical media \cite{Born} after the formulation of Maxwell's theory of electromagnetism in 1865 \cite{Maxwell}. The refractive index of an optical medium, of course, is considered after Maxwell's theory as the ratio of the phase speed of light (electromagnetic wave) to the speed of light in the free space \cite{Born}. Surprisingly, the precession measurement of the wavelength dependency of the refractive index of an optical medium is still of experimental interest \cite{Hansinger}. However, there had been an important progress in the determination of the relative refractive index (i.e. non-universal constant) in terms of the ratio of the phase speeds of light in the two media during the time duration between Snell's discovery (in 1621 \cite{Snell2}) and Rømer's discovery (in 1676 \cite{Roemer}). This progress was made by Descartes \cite{Descartes} and Fermat \cite{Fermat} by considering the speed of light to be finite in the optical media.

Descartes obtained pseudo-Snell law \cite{Descartes}
\begin{eqnarray}\label{eq1}
\frac{\sin(\theta_i)}{\sin(\theta_r)}=\frac{v_r}{v_i}
\end{eqnarray}
for the refraction of light from theoretical (metaphysical) point of view in 1637 by considering the conservation of the tangential\footnote{Here tangential means, tangential to the interface of the two media.} component of the velocity of a particle in the process of mechanical refraction and further considering a metaphysical form of the optical-mechanical analogy which now-a-days maps the group speed of light in the refracting medium ($v_r$) and that in the medium of incidence ($v_i$) to the respective particle velocities in the two media. The modern form of the optical-mechanical analogy, however, was proposed long after by Hamilton in 1834 \cite{Hamilton2}. The phase speed and group speed though are the same in a non-dispersive optical medium (free space), they are different in a dispersive optical medium and still is a subject of experimental study \cite{Danielmeyer,Hansinger,Babicz}. Descartes' derivation of the pseudo-Snell law was the first attempt for the determination of the relative refractive index ($n_{ri}=\frac{v_r}{v_i}=\frac{c_r}{c_i}$) in terms of the optical properties of the two non-dispersive media. The pseudo-Snell law, however, was rejected by Fermat in 1662 by introducing the principle of least ``time"\footnote{Here, the time is truly not least. Maupertuis' least abbreviated action principle ($\delta\int_{q_1}^{q_2}pdq=0$ \cite{Maupertuis,Landau}) can, however, be applied to a photon of (relativistic) energy $E=\hbar\omega$ in an optical medium of refracting index $n$. The magnitude of the (relativistic) momentum of the photon is given by $p=E/v_p$ where $v_p$ is the phase speed of the photon in the medium. Since the energy is conserved in the process of refraction, we can recast Maupertuis' least abbreviated action principle for the photon as $\delta\int_{q_1}^{q_2}\frac{dq}{v_p}=0$. The infinitesimal displacement of the photon ($dq$) over the infinitesimal time $dt$ is not same as $v_pdt$ because the phase speed does not represent the actual speed of a photon (particle) in a medium. The difference of these two, of course, was not known during Fermat's time. However, we can define $d``t"=\frac{dq}{v_p}$ as an infinitesimal abbreviated time. Thus we have $\delta\int_{q_1}^{q_2}d``t"=0$.  This is Fermat's least ``time" principle. By multiplying $c$ both the sides, it can be further recast as $\delta\int_{q_1}^{q_2}d``q"=0$ where $d``q"=\frac{c}{v_p}dq=ndq$ is an infinitesimal optical path-length. Thus the least``time" principle can also be called as the least optical path-length principle.} for light \cite{Fermat}. After the principle of least ``time", Snell's law of refraction takes the form \cite{Fermat}
\begin{eqnarray}\label{eq2}
\frac{\sin(\theta_i)}{\sin(\theta_r)}=\frac{c_i}{c_r}.
\end{eqnarray}
This form of Snell's law correctly determines the relative refractive index ($n_{ri}=\frac{c_i}{c_r}=\frac{c/c_r}{c/c_i}=\frac{n_r}{n_i}$) for a dispersive optical medium. The same form of Snell's law was also obtained later within Huygens' principle \cite{Huygens} and it takes the modern form of Snell's law ($n_r\sin(\theta_r)=n_i\sin(\theta_i)$ \cite{Born}). Eventuality, Descartes' derivation of the pseudo-Snell law is hardly given any importance now-a-days. However, Descartes' idea was brilliant. Unfortunately, the application of this idea went towards a wrong direction due to the non-existence of the special theory of relativity which, however, was introduced much later by Einstein in 1905 \cite{Einstein}. A careful extension of Descartes' method of deriving the pseudo-Snell law, however, can still be useful for the derivation of the refractive index within a relativistic treatment not only for the mechanical refraction but also for the optical refraction. This work was not done so far though an attempt in arriving at Eqns. (\ref{eq1}) and (\ref{eq2}) was made with the consideration of the conservations of the tangential component of the momentum and energy of a particle and further consideration of the dressing of the mass of the particle in the process of the mechanical refraction in the line of thoughts of aether drag \cite{Joyce}. We are overruling the idea of the dressing of the mass in the refracting media as the existence of aether has been made redundant after the introduction of the special theory of relativity \cite{Einstein}. Snell's law for the optical refraction has been obtained by Drosdoff-Widom within a particle point of view by considering conservations of the tangential component of the momentum and energy of a  photon \cite{Drosdoff}. Maupertuis' least abbreviated action principle \cite{Maupertuis,Landau} has recently been used by Luca \textit{et al} to arrive only at Eqns. (\ref{eq1}) and (\ref{eq2}) for the mechanical refraction of a particle \cite{Luca}. They did not, however, determine the refractive index for all possible values of the speed of the particle. We dedicate this article for a careful extension of Descartes' method to derive the refractive index for the mechanical refraction of a relativistic particle having an arbitrary speed and reach both the results Eqn. (\ref{eq1}) and Eqn. (\ref{eq2}) at the non-relativistic and ultra-relativistic limits, respectively.

Our article begins with the optical-mechanical analogy for a relativistic particle. In connection with the same we consider the analogy of Helmholtz's wave equation for light in an optical medium and the time-independent Klein-Gordon equation for a particle in a potential. We consider the conservation of the tangential component of the relativistic momentum and relativistic energy in the process of mechanical refraction of a relativistic particle. We define the refractive index for the mechanical refraction according to the optical-mechanical analogy from the relativistic point of view for all possible values of the speed of the particle. We show that the ultra-relativistic limit of the mechanical refractive index obtained by us matches with the form of Fermat's result on Snell's law \cite{Fermat} and the non-relativistic limit of the same matches with the form of Descartes' metaphysical result on pseudo-Snell law \cite{Descartes}.

\section{Refractive index for mechanical refraction}
\begin{figure}
\includegraphics[height=8cm,width=8cm]{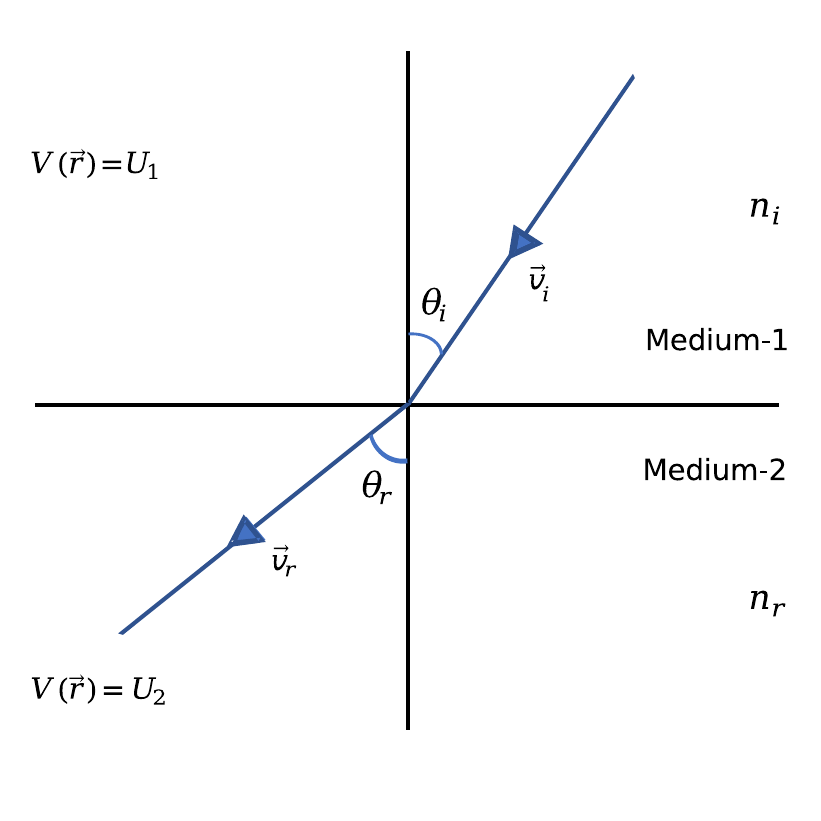}
\caption{Schematic diagram for mechanical refraction from medium-1 to medium-2 for $U_2>U_1$.}
\label{fig1}
\end{figure}

The time-independent form of the scalar wave equation, called as Helmholtz's wave equation, for the propagation of light in a medium of refractive index $n$ is given by \cite{Born2}
\begin{eqnarray}\label{eq3}
\bigg[\nabla^{2}+\frac{n^2\omega^2}{c^2}\bigg]\psi(\vec{r})=0
\end{eqnarray}
where $\omega$ is the (angular) frequency of the scalar field $\psi(\vec{r})$ which is the spatial part of one of the components of the electromagnetic field at the position $\vec{r}$ in the optical medium. Since the energy-momentum relation for a relativistic particle of energy $E$, momentum $\vec{p}$, and rest mass $m_0$ in a potential $V(\vec{r})$ takes the form $(E-V)^2=p^2c^2+m_0c^2$, it follows the  time-independent Klein-Gordon equation for the potential $V$ as \cite{Klein-Gordon}
\begin{eqnarray}\label{eq4}
\bigg[\nabla^{2}+\frac{[E-V(\vec{r})]^2-m_0^2c^4}{\hbar^2c^2}\bigg]\psi(\vec{r}) = 0
\end{eqnarray}
where $\psi(\vec{r})$ is the spatial part of the wave function at the position $\vec{r}$ for the case of spin 0\footnote{With spin 0, we are considering the particle to be a scalar particle.}. While the refraction of light at the interface of two optical media indicates a sudden change in the wave-vector ($\vec{k}$), the refraction of a particle at the interface of two media of different potential energy would indicate a sudden change in relativistic momentum ($\vec{p}$) according to the optical-mechanical analogy \cite{Hamilton2}. Since the potential energy is time-independent, the relativistic energy of the particle would be conserved in the process of the mechanical refraction. Thus the sudden change in the relativistic momentum would be caused by a finite discontinuity of the potential energy across the interface. We consider the case of two constant potentials across the interface as shown in figure \ref{fig1} so that there would be no potential gradient towards the direction tangential to the interface. This causes the conservation of the component of the relativistic momentum tangential to the interface. 

However, according to the optical-mechanical analogy \cite{Hamilton2,Cohen-Tannoudji,Dutta-Gupta}, the wave function for the free particle in the medium with either of the potential energy ($V(\vec{r})=U_1$ or $V(\vec{r})=U_2$) follows Helmholtz's wave equation 
\begin{eqnarray}\label{eq5}
\bigg[\nabla^{2}+\frac{n^2[E^2-m_0^2c^4]}{c^2\hbar^2}\bigg]\psi(\vec{r})=0
\end{eqnarray}
with the magnitude of the momentum of a photon ($\hbar\omega/c$) in the free space is replaced with the magnitude of the relativistic momentum ($+\sqrt{E^2-m_0^2c^4}$) of the particle in the free space ($V(\vec{r})=0$). Now comparing Eqns. (\ref{eq4}) and (\ref{eq5}), we get the refractive index for the mechanical refraction as
\begin{eqnarray}\label{eq6}
n(\vec{r})&=&+\frac{\sqrt{[E-V(\vec{r})]^2-m_0^2c^4}}{\sqrt{E^2-m_0^2c^4}}\nonumber\\&=&+\frac{\sqrt{(E-V(\vec{r})-m_0c^2)(E-V(\vec{r})+m_0c^2)}}{\sqrt{E^2-m_0^2c^4}}~~
\end{eqnarray}
where $V(\vec{r})=U_1$ $\forall~\vec{r}\in$ medium-1 (medium of incidence) and $V(\vec{r})=U_2$ $\forall~\vec{r}\in$ medium-2 (medium of refraction).

Eqn. (\ref{eq6}) is our desired result for the refractive index for the mechanical refraction. This can be called as mechanical refractive index. It is interesting to note that the mechanical refractive index of the medium, as defined in Eqn. (\ref{eq6}), also takes the simpler form $n(\vec{r})=+\frac{pc}{\sqrt{E^2-m_0^2c^4}}$ in terms of the magnitude of the relativistic momentum of the particle in the medium. It is clear from Eqn. (\ref{eq6}) that the mechanical refractive index decreases if the potential energy increases and consequently, the magnitude of the relativistic momentum of the particle also decreases. This means that when moving from free space into a medium with positive potential energy, the angle of refraction is greater than the angle of incidence as depicted in figure \ref{fig1}. Such a decrement of the refractive index corresponding to the increment of the potential energy of a particle is a well-known phenomenon in neutron interferometry \cite{Rauch}.

\begin{figure}
\includegraphics[height=6cm,width=8cm]{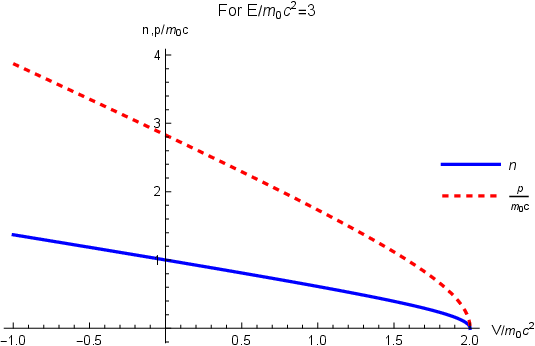}
\caption{The solid line indicates the plot of the refractive index with respect to the potential energy for mechanical refraction and the dashed line indicates the plot of the magnitude of the relativistic momentum with respect to the potential energy. While the solid line follows Eqn. (\ref{eq6}), the dashed line follows the relativistic energy-momentum relation $p=+\sqrt{(E-V)^2-m_0^2c^4}$.}
\label{fig2}
\end{figure}

The result obtained in Eqn. (\ref{eq6}) though is a relativistic one, it is a purely classical result as because the optical-mechanical analogy is valid only in the domains of the geometrical optics (wavelength $\lambda\rightarrow0$) and the classical mechanics ($\hbar\rightarrow0$). The time-independent Klein-Gordon equation and Helmholtz's wave equation have been used just to define the refractive index (in the limiting case of $\hbar\rightarrow0$) not to investigate any quantum mechanical effect. The refractive index as obtained in Eqn. (\ref{eq6}), however, is nothing but the ratio of the magnitude of relativistic momentum of a particle in a medium to the magnitude of the relativistic momentum of the particle in the free space ($V(\vec{r})=0$). We plot Eqn. (\ref{eq6}) in figure \ref{fig2} to show the potential energy dependence of the refractive index (solid line) for the mechanical refraction. We also plot the potential energy dependence of the relativistic momentum of the particle (dashed line) in the same figure. It is clear from figure \ref{fig2} that both the mechanical refractive index and the relativistic momentum reach $0$ as $E-V-m_0c^2\rightarrow0$. This is the non-relativistic limit for the refractive index. 

 The refractive index $n$ can be greater than $1$, as shown in figure \ref{fig2}, if the potential energy $V$ is negative which is possible for a particle in a potential well. Eventually, at the ultra-relativistic limit ($p/m_0c\rightarrow\infty$) which would be achieved by sending $V\rightarrow-\infty$ keeping $E$ fixed, the refractive index tends to infinity as indicated in figure \ref{fig2}.

In the non-relativistic regime ($p/m_0c\ll1$) Eqn. (\ref{eq6}), however, takes the form
\begin{eqnarray}\label{eq7}
n(\vec{r})\simeq+\frac{1}{\sqrt{E^2-m_0^2c^4}}\sqrt{2m_0c^2[E_{nr}-V(\vec{r})]}
\end{eqnarray}
where $E_{nr}=\frac{1}{2}m_0v^2+V(\vec{r})$ is the non-relativistic energy of the particle and $v$ is the speed of the particle. It should be mentioned in this regard that a similar form of the refractive index ($n(x)=\frac{1}{\hbar\omega}\sqrt{2m_0c^2[E_{nr}-V(x)]}$\footnote{Here $E_{nr}=\frac{\hbar^2[\vec{k}\cdot{\hat{x}}]^2}{2m_0}$ and $V(x)=-\frac{k^2}{2m_0}[n^2(x)-1]$.}) for light was obtained by Kay-Moses recasting Helmholtz's wave equation in the form of  time-independent Schrödinger equation to determine the reflection and transmission coefficients in the language of quantum mechanics for the 1-D scattering of light (of wave vector $\vec{k}=\frac{\omega n}{c}\hat{k}$) in an engineered dielectric medium of refractive index $n=n(x)$ \cite{Kay-Moses}. This form of the refractive index is still relevant for the theoretical \cite{Yang, Horsley} and experimental \cite{Szameit} study of the optical tunnelling of an evanescent wave in an engineered dielectric medium \cite{Cohen-Tannoudji, Dutta-Gupta}. However, the non-relativistic form of the refractive index obtained in Eqn. (\ref{eq7}) is not useful for a highly dispersive medium where the speed of the particle (as well as the group speed of a wave-packet) is not close to $c$. For such a case, one needs the relativistic form of the refractive index as obtained in Eqn. (\ref{eq6}).

\subsection{Law of mechanical refraction}

The mechanical refraction as described in figure \ref{fig1} must follow the conservation of relativistic energy and the tangential component of relativistic momentum as mentioned above. Let us now explore the conservation of these two physical quantities to find the law of mechanical refraction. For the mechanical refraction too, we take the angle of incidence to be as $\theta_i$ and the angle of refraction to be as $\theta_r$ as described in figure \ref{fig1}. While the magnitude of the relativistic momentum ($\vec{p}_i$) of the scalar particle of rest mass $m_0$ and relativistic energy $E$ in the medium of incidence (medium-1) is $p_i=+\frac{1}{c}\sqrt{(E-U_1)^2-m_0^2c^4}$, the magnitude of the relativistic momentum ($\vec{p}_r$) of the same particle in the medium of refraction (medium-2) is $p_r=+\frac{1}{c}\sqrt{(E-U_2)^2-m_0^2c^4}$ according to the conservation of the relativistic energy. Now, according to the conservation of the component of the relativistic momenta tangential to the interface ($p_i\sin(\theta_i)=p_r\sin(\theta_r)=constant$), we have
\begin{eqnarray}\label{eq8}
\frac{\sqrt{(E-U_1)^2-m_0^2c^4}}{c}\sin(\theta_i)=\frac{\sqrt{(E-U_2)^2-m_0^2c^4}}{c}\sin(\theta_r)\nonumber\\
\end{eqnarray}
in the process of the mechanical refraction as described in figure \ref{fig1}. By multiplying the denominators of both the sides of above equation by the constant $\sqrt{E^2-m_0^2c^4}/c$ we can recast it to the form
\begin{eqnarray}\label{eq9}
n_i\sin(\theta_i)=n_r\sin(\theta_r)
\end{eqnarray}
where $n_i=+\frac{\sqrt{[E-U_1]^2-m_0^2c^4}}{\sqrt{E^2-m_0^2c^4}}$ and $n_r=+\frac{\sqrt{[E-U_2]^2-m_0^2c^4}}{\sqrt{E^2-m_0^2c^4}}$ are the refractive indices of medium-1 and medium-2, respectively, and follow Eqn. (\ref{eq6}) for the mechanical refraction. Eqn. (\ref{eq9}) can be called as the law of mechanical refraction. 

\subsection{Law of mechanical refraction at ultra-relativistic and non-relativistic limits}
The velocity of the particle, can however, be determined from the Hamiltonian $H=+\sqrt{p^2c^2+m_0^2c^4}+V(\vec{r})$. The velocity of the particle can be determined from this Hamiltonian as $\vec{v}=\dot{\vec{r}}=\frac{\partial H}{\partial\vec{p}}=\frac{\vec{p}c}{\sqrt{p^2+m_0^2c^2}}$. From this expression we can easily show $\vec{p}=\frac{m_0\vec{v}}{\sqrt{1-v^2/c^2}}$ for all values of $v$ and $v\rightarrow c$ in the limiting case of $p/m_0c\rightarrow\infty$. Such a limiting case is referred to as the ultra-relativistic limit. Now we need to define the phase speed of the particle, according to the optical-mechanical analogy, as $v_p=\frac{H}{p}=c^2/v+[V(\vec{r})\sqrt{1-v^2/c^2}]/m_0v$. The group speed, on the other hand, would be the speed of the particle $v=\frac{pc}{\sqrt{p^2c^2+m_0^2c^2}}$ according to the optical-mechanical analogy. We can define the non-dispersive medium for the mechanical refraction, according to the optical-mechanical analogy, as a medium with the potential energy $V(\vec{r})$ such that $v_p=v$. This is possible only if $V(\vec{r})=-m_0c^2\sqrt{1-v^2/c^2}$. For the refraction of a mass-less particle (e.g. a photon) in a non-dispersive medium, we have $V(\vec{r})=0$. However, In the case of the ultra-relativistic limit, the phase speed of the particle takes the form $v_p\rightarrow c+\frac{c}{E/V(\vec{r})-1}$. It is interesting to note that the mechanical refractive index of the medium $n(\vec{r})=+\frac{pc}{\sqrt{E^2-m_0^2c^4}}$, as obtained in Eqn. (\ref{eq6}), takes the form $n\rightarrow\frac{c}{v_p}$ only at the ultra-relativistic limit.

\subsubsection{Snell's law}
The mechanical refractive index of the medium-1 can be written from Eqn. (\ref{eq6}) as $n_i=+\frac{p_ic}{\sqrt{E^2-m_0^2c^4}}$ where $p_i$ is the magnitude of the relativistic momentum of the particle in the medium-1.  At the ultra-relativistic limit ($p_i/m_0c\rightarrow\infty$), this expression of the mechanical refractive index takes the form $n_i\rightarrow\frac{c}{c_i}$ where $c_i=c\frac{E}{E-U_1}$ is the phase speed of the particle at the ultra-relativistic limit. Similarly,  from Eqn. (\ref{eq6}) we have the mechanical refractive index in medium-2 as $n_r=+\frac{p_rc}{\sqrt{E^2-m_0^2c^4}}$ where $p_r$ is the magnitude of the relativistic momentum of the particle in the medium-2. At the ultra-relativistic limit ($p_r/m_0c\rightarrow\infty$), it takes the form $n_r\rightarrow\frac{c}{c_r}$ where $c_r=c\frac{E}{E-U_2}$ is the phase speed of the particle in medium-2. Thus combining the above two results for the particle refractive indices of the two media, we get the law of mechanical refraction from Eqn. (\ref{eq9}) at ultra-relativistic limit ($p/m_0c\rightarrow\infty$) as
\begin{eqnarray}\label{eq10}
\frac{\sin(\theta_i)}{\sin(\theta_r)}\rightarrow\frac{c_i}{c_r}.
\end{eqnarray}
This ultra-relativistic result for the mechanical refraction takes the form of Fermat's result on Snell's law \cite{Fermat} as mentioned in Eqn. (\ref{eq2}). 

In reality, a photon having zero rest mass ($m_0\rightarrow0$) is an ultra-relativistic particle though its relativistic momentum is finite ($p=n\hbar\omega/c$). So, a photon with a single spin component must follow  Eqn. (\ref{eq10}) as expected from Snell's law for the refraction of light. However, it does not necessarily mean that the group speed of light (electromagnetic waves) has to take the form $v\rightarrow c$, like that for an ordinary particle with $m_0\neq0$, in order to reach the ultra-relativistic limit. The wavelength of the wave associated with the photon, however, takes the form $\lambda\sim 2\pi\hbar/p$. Incidentally, the wavelength takes a non-zero finite value which takes the case away from the geometrical optics limit where the optical-mechanical analogy does not hold. Hence, the group speed of light is not necessarily same as the speed ($v$) of a photon rather the signal speed of light is same as the speed of a photon in an optical medium \cite{Sommerfeld}. In order to hold the optical-mechanical analogy for a photon, we need to consider $p\rightarrow\infty$. In such a case, the group speed of light takes the form $v\rightarrow c$ even in a dispersive medium.

\subsubsection{Pseudo-Snell law}
On the other hand, at the non-relativistic limit ($p_i/m_0c\rightarrow0$), the mechanical refractive index of medium-1 as expressed in  Eqn. (\ref{eq6}) takes the form $n_i=\frac{p_ic}{\sqrt{E^2-m_0^2c^4}}\rightarrow\frac{v_i}{\sqrt{E^2-m_0^2c^4}/m_0c}$ where $v_i$ is the speed of the particle in medium-1. Here we are not taking the non-relativistic limit of $\sqrt{E^2-m_0^2c^4}$ as it is a constant in the process of refraction.  Similarly, we have the mechanical refractive index of medium-2 at the non-relativistic limit ($p_r/m_0c\rightarrow0$) as $n_r=\frac{p_rc}{\sqrt{E_0^2-m_0^2c^4}}\rightarrow\frac{v_r}{\sqrt{E^2-m_0^2c^4}/m_0c}$ where $v_r$ is the speed of the particle in medium-2. Thus combining the above two non-relativistic results for the two media we get the law of mechanical refraction from Eqn. (\ref{eq9}) as
\begin{eqnarray}\label{eq11}
\frac{\sin(\theta_i)}{\sin(\theta_r)}\rightarrow\frac{v_r}{v_i}.
\end{eqnarray}
This non-relativistic result for the mechanical refraction takes the form of Descartes' metaphysical result on pseudo-Snell law \cite{Descartes} as mentioned in Eqn. (\ref{eq1}). 

\section{Conclusion}
To conclude, we have analytically determined the mechanical refractive index for the refraction of a particle from the relativistic point of view. Eqn. (\ref{eq6}) is our key result for the same. A few other interesting results for the law of mechanical refraction are obtained in Eqns. (\ref{eq9}), (\ref{eq11}), and (\ref{eq10}). We have employed the optical-mechanical analogy with the correspondence of Helmholtz's wave equation for the propagation of light in an optical medium to the time-independent Klein-Gordon equation for the motion of a particle in a potential for the derivation of the mechanical refractive index. The mechanical refractive index obtained by us though is relativistic, it is purely a classical result. 

We have considered a scalar particle (spin 0) for the derivation of the mechanical refractive index. A photon, however, is a vector particle (spin 1). Since the refractive index of an optical medium is a property of geometrical optics which is analogous to classical mechanics, any particle with a specific component of spin will also have the same mechanical refractive index. Similarly, a photon with a specific component of spin will also have the same refractive index in the limiting case of $p/m_0c\rightarrow\infty$.

The optical-mechanical analogy with Helmholtz's wave equation and Klein-Gordon equation also mathematically allows us to take the negative square root of $[E-V(\vec{r})]^2-m_0^2c^4$ in Eqn. (\ref{eq6}) to get a negative refractive index $n(\vec{r})=-\frac{\sqrt{[E-V(\vec{r})]^2-m_0^2c^4}}{\sqrt{E^2-m_0^2c^4}}$ for the mechanical refraction. In reality, the negative refractive index is a property of a metamaterial \cite{Bose} under an optical refraction for a certain frequency range ($\omega\sim$ GHz - THz) for which both the permittivity and the permeability of the medium are negative \cite{Veselago,Shelby}. Such a property, however, is not seen for a very large frequency. The optical-mechanical analogy we have adopted here is applicable for a very large frequency (ideally $\omega\rightarrow\infty$). This analogy, however, does not allow any one-to-one correspondence of the negative permittivity and the negative permeability to the mechanical quantities ($E,V(\vec{r})$, and $m_0$). Thus we don't have an analogy of the negative mechanical refractive index with the negative optical refractive index.     

The motion of the particle in the process of refraction though follows Hamilton's least action principle ($\delta\int_{t_1}^{t_2}L(q,\dot{q},t)=0$ \cite{Hamilton3, Landau2, Rojo}) when the time is taken into account for the path of the particle and Maupertuis' least abbreviated action principle  ($\delta\int_{q_1}^{q_2}pdq=0$ \cite{Maupertuis,Landau}) when the time is not taken into account for the trajectory of the particle \cite{Landau}, the refraction is a consequence of the conservation of the relativistic energy and the tangential component of the relativistic momentum. The conservation of the tangential component of the wave vector of the electromagnetic field (which is analogous to the momentum of a photon) is also used as boundary conditions to obtain Snell's law from Maxwell's equations \cite{Maxwell} for the optical refraction across two linear media with no free charge and free current at the interface in the limiting case of the geometrical optics \cite{Born,Griffiths}. In general, the normal component of the electric field and the tangential component of the magnetic field are discontinuous at the interface for an optical refraction \cite{Griffiths}. The normal component of the electric displacement vector, however, is continuous for the optical refraction across two linear dielectric media \cite{Griffiths}.

In optics, we usually work with angular frequency $\omega$ and the wave-vector $\vec{k}$ of an electromagnetic wave rather than the energy ($E=\hbar\omega$) and momentum ($\vec{p}=\hbar\vec{k}$) of a photon. While the angular frequency of the incident electromagnetic wave is unaltered in the process of optical refraction, the relativistic energy of the particle is conserved in the process of the mechanical refraction. The wavenumber ($k=n\omega/c$) and the magnitude of the relativistic momentum, on the other hand, on the other hand, are not conserved in the process of the optical refraction and the mechanical refraction, respectively. However, the electromagnetic field also has a momentum called either Minkowski momentum ($\vec{p}_{\text{Min}}=\int\text{d}^3\vec{r}\vec{D}\times\vec{B}$) or Abraham momentum ($\vec{p}_{\text{Abr}}=\int\text{d}^3\vec{r}\frac{\vec{E}\times\vec{H}}{c^2}$) \cite{Barnett}. On the other hand, the medium also has a momentum due to the coupling of the electromagnetic field with the electric and magnetic dipoles in it. While the Abraham momentum is of kinetic momentum type, the Minkowski momentum is of canonical momentum type. For a photon, while the canonical momentum in a medium takes the form $\vec{p}=\frac{n\hbar\omega}{c}\hat{k}$, the kinetic momentum in the same medium takes the form $\vec{p}_{\text{kin}}=\frac{\hbar\omega}{cn}\hat{k}$. The total (kinetic or canonical) momentum i.e. the field momentum and the momentum of the medium, however, is a conserved quantity in the process of the optical refraction \cite{Barnett}.
We, of course, don't have such a vector field description for the mechanical refraction of a scalar particle.

For a photon with a single spin component in a non-dispersive medium, the least action principle, however, becomes the least ``time" principle \cite{Luca}. Hence Fermat's theory of the least ``time" \cite{Fermat} is a relativistic theory though the actual 4-vector formulation of the theory of relativity \cite{Einstein,Resnick} was done long after his death. Huygens-Fresnel principle \cite{Huygens,Fresnel,Born3}, of course, goes one step further for the wave optics. All the theories of light which are found correct, are by default relativistic theory as because photon is an ultra-relativistic particle. So, the derivation of Snell's law of refraction needs a relativistic treatment. 

Descartes' metaphysical theory for the refraction of light was essentially a non-relativistic theory and was inconsistent with the phase speed of light in a denser or rarer medium \cite{Descartes}. Incidentally this metaphysical theory was rejected by Fermat introducing the principle of least ``time" \cite{Fermat}. Since then Descartes' metaphysical theory of the refractive index has not been respected well. Descartes considered the conservation of the component of the ordinary velocity (of both the particle and light) tangential to the interface of the two media for the process of refraction \cite{Descartes}.  This idea, however, was brilliant. We have extended this idea and considered the conservation of the tangential component of the relativistic momentum along with the relativistic energy for the derivation of the mechanical refractive index. While the ultra-relativistic limit of the mechanical refractive index obtained by us matches with the form of Fermat's result on Snell's law \cite{Fermat}, the non-relativistic limit of the same matches with the form of Descartes' metaphysical result on pseudo-Snell law \cite{Descartes}.

The mechanical refractive index obtained by us may find potential applications towards the transmission and reflection of particles by engineered (mechanical) media in analogy with the optical transmission and reflection by engineered dielectric media as mentioned in the introductory section. Semiclassical study of the transmission and reflection coefficients for a particle in an engineered media can be studied both in non-relativistic regime and ultra-relativistic regime by using the mechanical refractive index (Eqn. (\ref{eq6})) obtained by us. A classical path integral formulation was considered to determine the refractive index of an optical medium in terms of multiple scattering of a photon due to the atoms/molecules in the medium  \cite{Grooth}. A quantum mechanical study \cite{Mead} of the same with the path integral approach (beyond the optical-mechanical analogy) for the mechanical refractive index is kept as an open problem.

\subsection*{Acknowledgement} 
S. Biswas acknowledges partial financial support of the SERB, DST, Govt. of India under the EMEQ Scheme [No. EEQ/2023/000788].


\begin{thebibliography}{99}
\bibitem{Snell1}R. Rashed, \textit{A pioneer in anaclastics: Ibn Sahl on burning mirrors and lenses}, \href{https://doi.org/10.1086/355456}{ISIS \textbf{81}, 464 (1990)}; A. Kwan, J. Dudley, and E. Lantz, \textit{Who really discovered Snell's law?}, \href{https://doi.org/10.1088/2058-7058/15/4/44}{Phys. World \textbf{15}, 64 (2002)}

\bibitem{Snell2}C. Huygens, \textit{Dioptrica}, ed. B. de Volder and B. Fullenius (\href{https://www.google.co.in/books/edition/Opuscula_postuma/tAsRG4yIJKwC?hl=en&gbpv=1&dq=Christiaan+Huygens,+Opuscula+Posthuma&printsec=frontcover}{Opuscula Posthuma}, Leiden, 1703)


\bibitem{Huygens}C. Huygens, \textit{Traité de la Lumière}, (Van der Aa, Leiden 1690); English Translation: S. P. Thompson, \textit{\href{https://archive.org/details/huyghens-traite-de-la-lumiere-gauthier-villars-1690-english-trans/mode/2up}{Treatise on Light}} (Macmillan, London, 1912)

\bibitem{Fresnel}A. Fresnel, \textit{Mémoire sur la diffraction de la lumière}, \href{https://www.biodiversitylibrary.org/item/55242#page/1/mode/1up}{Mémoires de l'Académie Royale des Sciences de l'Institut de France (Paris) \textbf{5}}, 339 (1826)

\bibitem{Roemer}O. Römer, \textit{Démonstration touchant le mouvement de la lumière trouvé par}, J. des Sçavans, 276 (Paris, 7 Dec. 1676); English Translation: \textit{A demonstration concerning the motion of light}, \href{https://doi.org/10.1098/rstl.1677.0024}{Philos. Trans. R. Soc. (London) \textbf{12}, 893 (1677)}

\bibitem{Newton}I. Newton, \textit{Opticks: or, A treatise of the reflexions, refractions, inflexions and colours of light. Also two treatises of the species and magnitude of curvilinear figures}, (\href{https://doi.org/10.5479/sil.302475.39088000644674}{Printed for S. Smith and B. Walford, London, 1704})

\bibitem{Young}T. Young, \textit{A course of lectures on natural philosophy and the mechanical arts}, vol. 1, lec. 35: on the theory of optics, \href{https://archive.org/details/lecturescourseof01younrich/page/412/mode/2up}{p. 413} (Printed for J. Johnson, London, 1807)

\bibitem{Fresnel2}A. Fresnel, \textit{Mémoire sur la Loi des modifications que la réflexion imprime à la lumière polarisée}, \href{https://www.biodiversitylibrary.org/item/55245#page/9/mode/1up}{Mémoires de l'Académie Royale des Sciences de l'Institut de France (Paris) \textbf{11}}, 393 (1832) 

\bibitem{Hamilton}W. R. Hamilton, \textit{Researches respecting vibration, connected with the theory of light}, \href{https://www.biodiversitylibrary.org/item/19854#page/1/mode/1up}{Proc. Royal Irish Academy \textbf{1}}, 341 (1841). See Ref. \cite{Rayleigh} for the group speed of sound waves.

\bibitem{Rayleigh}J. W. Strutt (B. Rayleigh), \textit{The Theory of Sound}, vol. 1, sec. 190 \& 191, pp. 244-249 (Macmillan, London, 1877); vol. 2, pp. 297-302 (Macmillan, London, 1878); The \href{https://doi.org/10.1017/CBO9781139058087}{vol. 1} and \href{https://doi.org/10.1017/CBO9781139058094}{vol. 2} are reprinted (Cambridge University Press, Cambridge, 2011).

\bibitem{Born}M. Born and E. Wolf, \href{https://doi.org/10.1017/CBO9781139644181}{\textit{Principles of Optics}}, 7th ed., sec. 1.5.1, pp. 38-40 and sec. 3.3.2, pp. 132-133   (Cambridge University Press, Cambridge, 1999)

\bibitem{Maxwell}J. C. Maxwell, \textit{A dynamical theory of the electromagnetic field}, \href{https://doi.org/10.1098/rstl.1865.0008}{Philos. Trans. R. Soc. (London) \textbf{155}, 459 (1865)}

\bibitem{Hansinger}P. Hansinger, P. Töpfer, N. Dimitrov, D. Adolph, D. Hoff, T. Rathje, A. M. Sayler, A. Dreischuh, and G. G. Paulus, \textit{Refractive index dispersion measurement using carrier-envelope phasemeters}, \href{https://doi.org/10.1088/1367-2630/aa5ca3}{New J. Phys. \textbf{19}, 023040 (2017)}

\bibitem{Descartes}R. Descartes, \textit{La Dioptrique}, Discours II: de la Réfraction (I. Maire, Leiden, 1637). Reprinted in \textit{Oeuvres de Descartes: Discours de la Méthode \& Essais} \href{https://www.biodiversitylibrary.org/item/88360#page/112/mode/1up}{VI}, ed. C. Adam and P. Tannery, pp. 93-105 (Léopold Cerf, Paris, 1902)

\bibitem{Fermat}P. de Fermat, \textit{A letter on réfraction to C. de la Chambre} (Paris, 1 Jan. 1662). Reprinted in \textit{Oeuvres de Fermat}, ed. P. Tannery and C. Henry, \href{https://www.biodiversitylibrary.org/item/62833#page/9/mode/1up}{vol. 2}, pp. 457-463 (Gauthier-Villars, Paris, 1894)

\bibitem{Hamilton2}W. R. Hamilton, \textit{On the application to dynamics of a general mathematical method previously applied to optics}, \href{https://www.biodiversitylibrary.org/item/46627#page/566/mode/1up}{British Association Report \textbf{4}\footnote{Here 4 means the 4th meeting.}, 513 (1834)}. A. S. Kompaneyets, \textit{A Course of Theoretical Physics}, \href{https://archive.org/details/KompaneyetsA.S.ACourseOfTheoreticalPhysicsVol.1FundamentalLawsMir1978/Kompaneyets}{vol. 1 (Fundamental Laws)}, sec. 21, pp. 269-278 (Mir, Moscow, 1978). Also see Ref. \cite{Hamilton3} for the same.

\bibitem{Danielmeyer}H. G. Danielmeyer and H. P. Weber, \textit{Direst measurement of the group velocity of light}, \href{https://doi.org/10.1103/PhysRevA.3.1708}{\pra{3}{1708}{1970}}

\bibitem{Babicz}M. Babicz, S. Bordoni, A. Fava, U. Kose, M. Nessi, F. Pietropaolo, G. L. Raselli, F. Resnati, M. Rossella, P. Sala, F. Stockera, and A. Zani, \textit{A measurement of the group velocity of scintillation light in liquid argon}, \href{https://doi.org/10.1088/1748-0221/15/09/P09009}{JINST \textbf{15}, P09009 (2020)}

\bibitem{Einstein}A. Einstein, \textit{Zur elektrodynamik bewegter körper}, \href{https://doi.org/10.1002/andp.2005517S113}{Annalen der Physik \textbf{17}, 891 (1905)}; \textit{Ist die trägheit eines körpers von seinem energieinhalt abhängig?}, \href{https://doi.org/10.1002/andp.19053231314}{Annalen der Physik \textbf{18}, 639 (1905)}

\bibitem{Joyce}W. B. Joyce and A. Joyce, \textit{Descartes, Newton, and Snell’s law}, \href{https://doi.org/10.1364/JOSA.66.000001}{J. Opt. Soc. Am. \textbf{66}, 1 (1976)}

\bibitem{Drosdoff}D. Drosdoff and A. Widom, \textit{Snell’s law from an elementary particle viewpoint}, \href{https://doi.org/10.1119/1.2000974}{Am. J. Phys. \textbf{73}, 973 (2005)}

\bibitem{Maupertuis}P. L. M. de Maupertuis, \textit{Accord de différentes loix de la nature qui avoient jusqu’ici paru incompatibles}, \href{https://www.biodiversitylibrary.org/item/55242#page/1/mode/1up}{Mémoires de l'Académie Royale des Sciences de l'Institut de France (Paris)}, 417 (15 Apr. 1744); Reprinted in \textit{Essay de Cosmologie}, pp. 154-173 (\href{https://gallica.bnf.fr/ark:/12148/bpt6k15100153/f178.item}{Mens Agitat Molem. Æneid. Lib. VI.} 1750)

\bibitem{Landau}L. D. Landau and E. M. Lifshitz, \text{Mechanics}, 3rd edn., sec. 44, pp. 140-143 (Butterworth-Heinemann, Oxford, 1976)

\bibitem{Luca}R. D. Luca, M. Di Mauro, A. Naddeo, \href{https://doi.org/10.1590/1806-9126-RBEF-2019-0339}{Revista Brasileira de Ensino de Física \textbf{42}, e20190339 (2020)}

\bibitem{Born2}See sec. 1.3, pp. 14-15 of Ref.\cite{Born}.

\bibitem{Klein-Gordon}O. Klein, \textit{Quantentheorie und fünfdimensionale relativitätstheorie}, \href{https://doi.org/10.1007/BF01397481}{Z. Physik \textbf{37}, 895 (1926)}; W. Gordon, \textit{Der comptoneffekt nach der Schrödingerschen theorie}, \href{https://doi.org/10.1007/BF01390840}{Z. Physik \textbf{40}, 117 (1926)}; V. Fock, \textit{Zur Schrödingerschen wellenmechanik}, \href{https://doi.org/10.1007/BF01399113 }{Z. Physik \textbf{38}, 242 (1926)}

\bibitem{Cohen-Tannoudji}C. Cohen-Tannoudji, B. Diu, and F. Laloë, \textit{Quantum Mechanics}, 2nd edn., vol. 1, ch. 1, sec. D-2-b \& D-2-c, pp. 27-30 (Wiley-VCH, Weinheim, 2020)

\bibitem{Dutta-Gupta}S. Dutta Gupta, A. Banerjee, and N. Ghosh, \textit{\href{https://doi.org/10.1201/b19330}{Wave Optics: Basic Concepts and Contemporary Trends}}, sec. 9.5, pp. 183-185 (Taylor \& Francis, New York, 2016)

\bibitem{Rauch}H. Rauch and S. A. Werner, \textit{\href{https://doi.org/10.1093/acprof:oso/9780198712510.001.0001}{Neutron Interferometry}}, 2nd edn, sec. 1.1, pp. 9-9 (Oxford University Press, Oxford, 2015)

\bibitem{Kay-Moses}I. Kay and H. E. Moses, \textit{Reflectionless transmission through dielectrics and scattering potentials},  \href{https://doi.org/10.1063/1.1722296}{J. Appl. Phys. 27, 1503 (1956)}

\bibitem{Yang}E. Yang, Y. Lu, Y. Wang, Y. Dai, and P. Wang, \textit{Unidirectional reflectionless phenomenon in periodic ternary layered material}, \href{https://doi.org/10.1364/OE.24.014311}{Opt. Express \textbf{24}, 14311 (2016)}

\bibitem{Horsley}S. A. R. Horsley and S. Longhi, \textit{Spatiotemporal deformations of reflectionless potentials},  \href{https://doi.org/10.1103/PhysRevA.96.023841}{\pra{96}{023841}{2017}}

\bibitem{Szameit}A. Szameit, F. Dreisow, M. Heinrich, S. Nolte, and A. A. Sukhorukov, \textit{Realization of reflectionless potentials in photonic lattices},  \href{http://dx.doi.org/10.1103/PhysRevLett.106.193903}{\prl{106}{193903}{2011}}

\bibitem{Sommerfeld}A. Sommerfeld, \textit{Über die fortpflanzung des lichtes in dispergierenden medien}, \href{https://doi.org/10.1002/andp.19143491002}{Annalen der Physik \textbf{44}, 177 (1914)}; L. Brillouin, \textit{Über die fortpflanzung des lichtes in dispergierenden medien}, \href{https://doi.org/10.1002/andp.19143491003}{Annalen der Physik \textbf{44}, 203 (1914)}; A. Sommerfeld, \textit{Optics (Lectures on Theoretical Physics Vol. IV)}, (Academic Press, New York, 1954); Also see sec. 22, pp. 114-135 of the Indian reprint of the book (Levant Books, Kolkata, 2006) 

\bibitem{Bose}J. C. Bose, \textit{On the rotation of plane of polarisation of electric waves by a twisted structure}, \href{https://doi.org/10.1098/rspl.1898.0019}{Proc. R. Soc. Lond. \textbf{63}, 146 (1898)}

\bibitem{Veselago}V. G. Veselago, \textit{The electrodynamics of substances with simultaneously negative values of $\epsilon$ and $\mu$}, \href{https://doi.org/10.1070/PU1968v010n04ABEH003699}{Sov. Phys. Usp. \textbf{10}, 509 (1967)}

\bibitem{Shelby}R. A. Shelby, D. R. Smith, and S. Shultz, \textit{Experimental verification of a negative index of refraction}, \href{https://doi.org/10.1126/science.1058847}{\sc{292}{77}{2001}}

\bibitem{Hamilton3}W. R. Hamilton, \textit{On a general method in dynamics; by which the study of the motions of all free systems of attracting or repelling points is reduced to the search and differentiation of one central relation, or characteristic function}, \href{https://www.jstor.org/stable/108066}{Philos. Trans. R. Soc. (London) \textbf{124}, 247 (1834)}; W. R. Hamilton, \textit{Second essay on a general method in dynamics}, \href{https://www.jstor.org/stable/108119}{Philos. Trans. R. Soc. (London) \textbf{125}, 95 (1835)}

\bibitem{Landau2}See sec. 2, pp. 2-4 of Ref.\cite{Landau}.

\bibitem{Rojo}A. Rojo and A. Bloch, \textit{\href{https://doi.org/10.1017/9781139021029}{The Principle of Least Action: History and Physics}}, (Cambridge University Press, Cambridge, 2018)

\bibitem{Griffiths}D. J. Griffiths, \textit{Introduction to Electrodynamics}, 4th Indian edn, sec. 9.3.1, pp. 407-407, (Pearson, Noida, 2015)

\bibitem{Barnett}S. M. Barnett, \textit{Resolution of the Abraham-Minkowski dilemma}, \href{https://doi.org/10.1103/PhysRevLett.104.070401}{\prl{104}{070401}{2010}}

\bibitem{Resnick}R. Resnick, \textit{Introduction to Special Relativity},  (Wiley, New York, (1968)

\bibitem{Born3}See sec. 8.2, pp. 413-417 of Ref.\cite{Born}.

\bibitem{Grooth}D. G. de Grooth, \textit{Why is the propagation velocity of a photon in a transparent medium reduced?} \href{https://doi.org/10.1119/1.18754}{Am. J. Phys. \textbf{65}, 1156 (1997)}

\bibitem{Mead}C. A. Mead, \textit{Quantum theory of the refractive index}, \href{https://doi.org/10.1103/PhysRev.110.359}{\prv{110}{359}{1958}}
\end{thebibliography}
\end{document}